\documentclass[a4paper]{jpconf}
\usepackage{graphicx}

\usepackage{amssymb,amsmath,array}
\usepackage{wrapfig,rotating}
\usepackage{xspace}
\usepackage{lineno}

\begin{document}
\title{Measurements of top-quark properties at the Tevatron}
\author{Andreas Werner Jung}

\address{Fermilab, MS 357, P.O. Box 500, Batavia, IL, 60510, USA\\
FERMILAB-CONF-13-044-PPD}

\ead{ajung@fnal.gov}


\newcommand{\dzero}     {D\O\xspace}
\newcommand{\wplus}     {$W+$jets\xspace}
\newcommand{\zplus}     {$Z+$jets\xspace}
\newcommand{\muplus}    {$\mu +$jets\xspace}
\newcommand{\eplus}     {$e +$jets\xspace}
\newcommand{\ljets}     {$\ell +$jets\xspace}
\newcommand{\ttbar}     {$t\bar{t}$\xspace}
\newcommand{\met}       {$\not\!\!E_T$\xspace}


\begin{abstract}
Recent measurements of top-quark properties at the Tevatron are presented. CDF uses data corresponding up to $9.0~\mathrm{fb^{-1}}$ to measure the ratio $R$ of the branching fractions ${\cal B}(t \rightarrow Wb) / {\cal B}(t \rightarrow Wq)$, the branching fraction for top-quarks decaying into $\tau$ leptons and the cross section for the production of an additional $\gamma$ in \ttbar production. The results from all these measurements agree  well with their respective Standard Model expectation. \dzero uses $5.3~\mathrm{fb^{-1}}$ of data to measure the \ttbar cross section as a function of the time. A time dependency would imply Lorentz invariance violation as implemented by the Standard Model extension. No time dependency is observed and \dzero sets first limits in the top-quark sector for Lorentz invariance violation. \dzero also determines indirectly the top quark width using the results of earlier measurements at \dzero. The measured top quark width is in agreement with the SM expectation and does not show any hints for new physics contributions.
\end{abstract}

\section{Introduction}
The top quark is the heaviest known elementary particle and was discovered at the Tevatron $p\bar{p}$ collider in 1995 by the CDF and \dzero collaboration \cite{top_disc1,top_disc2} with a mass around $173~\mathrm{GeV}$. The production is dominated by the $q\bar{q}$ annihilation process with 85\% as opposed to gluon-gluon fusion which contributes only 15\%. The top quark has a very short lifetime, which prevents any hadronization process of the top quark. Instead bare quark properties can be observed by measuring top quark properties.\\
The measurements presented here are performed using either the dilepton ($\ell \ell$) final state or the lepton+jets (\ljets) final state. Within the \ljets~final state one of the $W$ bosons (stemming from the decay of the $top$ quarks) decays leptonically, the other $W$ boson decays hadronically. For the dilepton final state both $W$ bosons decay leptonically. The branching fraction for top quarks decaying into $Wb$ is almost 100\%. Jets originating from a beauty quark ($b$-jets) are identified by means of a neural network (NN) built by the combination of variables describing the properties of secondary vertices and of tracks with large impact parameters relative to the primary vertex.

\section{Measurement of the ratio ${\mathbf R}$ of branching fractions (CDF)}
\begin{figure}[ht]
     \centerline{
       \includegraphics[width=0.975\columnwidth]{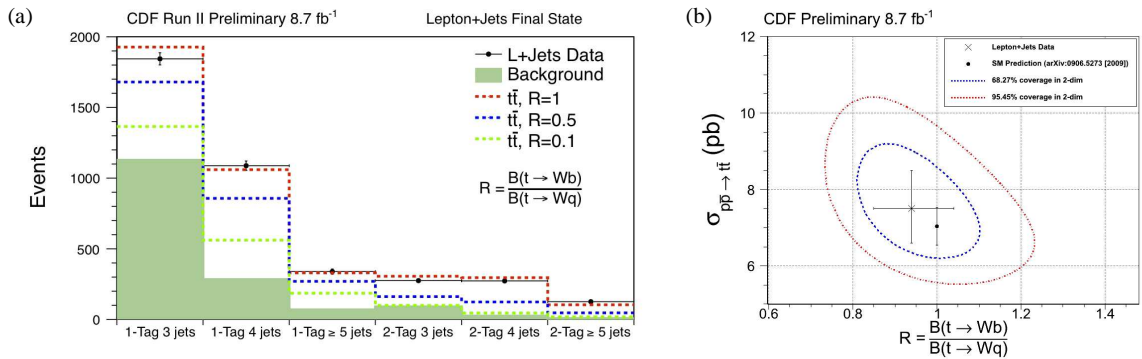}
     }
  \caption{\label{fig:CDF_R} (a) shows the number of data events with their respective number of background events for different jet and $b$-jet bins compared to three different $R$ templates: $R = 0.1$ (green dashed histogram), $R= 0.5$ (blue dashed histogram) and $R=1.0$ (red dashed histogram). (b) shows the result of a simultaneous log-likelihood fit to the \ttbar cross section and the ratio $R$ with the 2 and 3 standard deviation (s.d.) contours.}
 \end{figure}

CDF uses all available data corresponding to $8.7~\mathrm{fb^{-1}}$ in the \ljets decay channel to measure the ratio $R$ of the branching fractions ${\cal B}(t \rightarrow Wb) / {\cal B}(t \rightarrow Wq)$ \cite{cdf_R}. The data is selected by requiring  a lepton, missing transverse energy \met, at least three jets with exactly one or two $b$-jets. Figure \ref{fig:CDF_R}(a) shows the number of data events with their respective number of background events for different jet and $b$-jet bins. As an example three different templates with different values of $R$ are shown: $R = 0.1$ (green dashed histogram), $R= 0.5$ (blue dashed histogram) and $R=1.0$ (red dashed histogram).
A simultaneous log-likelihood fit to $R = {\cal B}(t \rightarrow Wb) / {\cal B}(t \rightarrow Wq)$ and to the \ttbar cross section $\sigma_{p\bar{p} \rightarrow t\bar{t}}$ is applied. The resulting log-likelihood is shown in Figure \ref{fig:CDF_R}b). An inclusive \ttbar cross section of $\sigma_{p\bar{p} \rightarrow t\bar{t}} = 7.5 \pm 1.0~(\mathrm{stat.+syst.})$ is measured in good agreement with earlier measurements at the Tevatron. The ratio $R$ of the branching fractions is measured to $R = 0.94 \pm 0.09~(\mathrm{stat.+syst.})$. The presented measurement of $R$ agrees with the SM expected value and with measurements from \dzero and CMS: an earlier measurement of $R$ by \dzero results in $R = 0.90 \pm 0.04 (\mathrm{stat.+syst.})$ \cite{d0_R} and the currently most precise $R$ measurement from CMS presented at this conference yields $R = 0.98 \pm 0.04$ \cite{CMS_R}. Assuming 3 generations and unitarity of the CKM matrix CDF also measured $V_{tb}$ to $0.97 \pm 0.05 (\mathrm{stat.+syst.})$, which is in good agreement with the SM expectation.\\

\section{Measurement of the branching fraction to ${\mathbf \tau}$ leptons (CDF)}
\begin{figure}[ht]
     \centerline{
       \includegraphics[width=0.785\columnwidth]{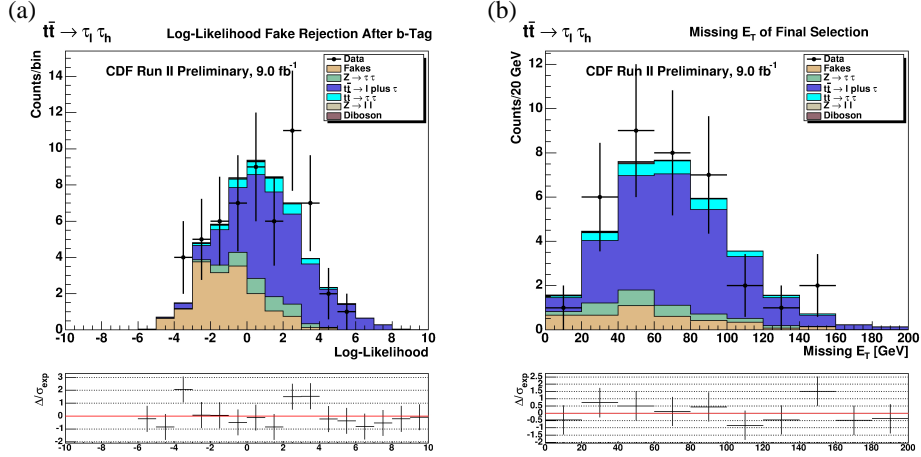}
     }
  \caption{\label{fig:CDF_tau} (a) shows the log-likelihood distribution before applying any cut, whereas (b) shows the final event selection after applying the log-likelihood cut of $0$.}
\end{figure}

The measurement of the branching fraction for top-quarks decaying into $\tau$ leptons ${\cal B}(t \rightarrow Wb \rightarrow \tau \nu_{\tau} b)$ is measured by CDF using $9.0~\mathrm{fb^{-1}}$ of data in the dilepton decay channel \cite{cdf_tau}. It relies on the precise measurement of the $\tau$ cross section in \ttbar production, which selects an $e$ or $\mu$ originating from the $W$ decay or the leptonic $\tau$ decay and thus allows for one hadronic $\tau$ decay (more details can be found in \cite{tau_gianluca}). In addition to the usual kinematic cuts an requirements on identified $b$-jets events are selected by imposing a certain log-likelihood cut enhancing $\tau$ fraction of events. The likelihood combines the following variables: \met, transverse mass of lepton plus \met and the transverse energy of the third highest $E_T$ jet. Figure \ref{fig:CDF_tau}(a) shows the log-likelihood distribution before applying any cut.
Imposing a minimum log-likelihood value of $0$ yields the final event selection with 36 observed events to be compared with the SM expectation of $33.6 \pm 0.7 ^{+3.7}_{-3.8}$ events. Figure \ref{fig:CDF_tau}(b) shows the final event selection after applying the log-likelihood cut of $0$; data are nicely described by the signal and background contributions. A branching fraction from decays to single and di-$\tau$ events of ${\cal B}(t \rightarrow Wb \rightarrow \tau \nu_{\tau} b) = 0.120 \pm 0.030 (\mathrm{stat.}) ^{+0.022}_{-0.019} (\mathrm{syst.}) \pm 0.007 (\mathrm{lumi})$ is measured. A further tightened likelihood cut using additional information from the transverse energy of the lepton $E_T(\ell)$ and $\Delta \phi (\not\!\!E_T,\ell)$ allows for a measurement of the branching fraction using only di-$\tau$ events: $0.098 \pm 0.022 (\mathrm{stat.}) \pm 0.014 (\mathrm{syst.})$. The results from these measurements agree well with the Standard Model expectation of ${\cal B}(t \rightarrow Wb \rightarrow \ell \nu_{\ell} b) = 0.108 \pm 0.009$ (averaged over $e$, $\mu$ and $\tau$ decay modes.

\section{Measurement of the $\mathbf{t \bar{t} + \gamma}$ production cross section (CDF)}
\begin{figure}[ht]
     \centerline{
       \includegraphics[width=0.425\columnwidth]{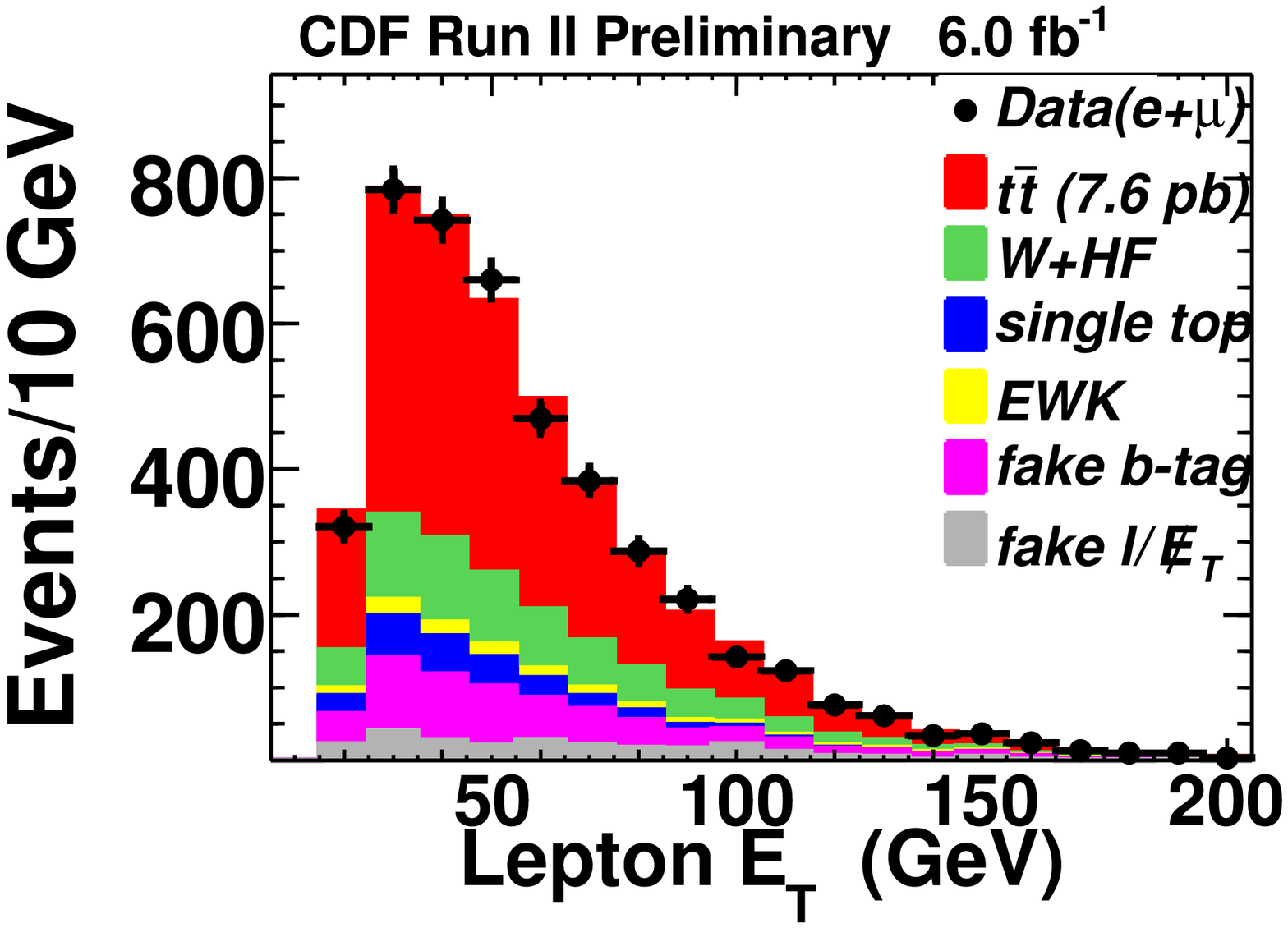}
       \includegraphics[width=0.425\columnwidth]{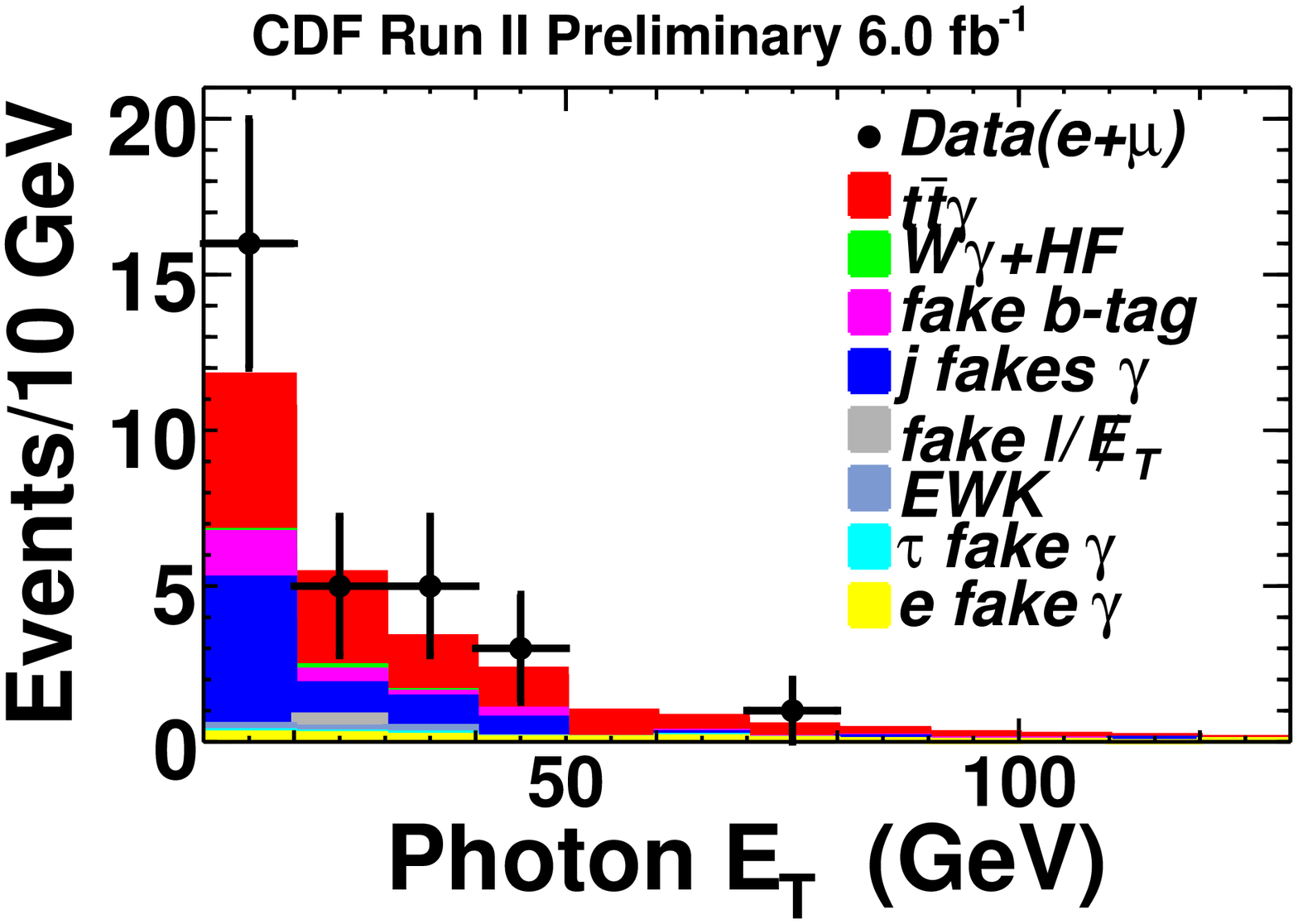}
     }
  \caption{\label{fig:CDF_photon} (left) shows that pre-selected data events compared to the various background contributions, whereas (right) shows the \ttbar $+ \gamma$ contribution after requiring a tight photon identification criteria.}
 \end{figure}

In addition to determining decay branching fractions CDF also measures the cross section for the production of an additional particle in \ttbar production: $p\bar{p} \rightarrow t\bar{t} + \gamma$ \cite{cdf_photon}. The measurement is based on the \ljets decay channel and thus events are selected by requiring a lepton, \met, at least three jets with one identified $b$-jet and a photon. The photon candidate is required to have $E_T^{\gamma} > 10$ GeV and no track with $p_T > 1$ GeV and at most one
track with $p_T < 1$ GeV, pointing at the calorimeter cluster; and minimal leakage into the hadronic calorimeter. Figure \ref{fig:CDF_photon} (left) shows that pre-selected data events are well described by all the various background contributions.
A tight $\chi^2$ cut on the photon identification suppresses largely the background and the signal contribution (red) is clearly visible in Figure \ref{fig:CDF_photon} (right). The most dominant remaining background contribution originates from miss-identification of jets as photons. In order to get a handle on this contribution the isolation cuts are not applied and $Z \rightarrow ee$ events are used to extrapolate the isolation shape to the used cuts at higher isolation values. Taking into account this fake photon contribution, all the other backgrounds and the signal contribution the expected amount of events is $26.9 \pm 3.4$ ($e$ and $\mu$ channel combined). The probability of the background alone to mimic the observed signal of 30 events is: 3 standard deviations. The cross section for $p\bar{p} \rightarrow t\bar{t} + \gamma$ is measured to $0.18 \pm 0.07 (\mathrm{stat.}) \pm 0.04 (\mathrm{syst.}) \pm 0.01 (\mathrm{lumi})$, which is a factor of 40 lower than the inclusive \ttbar cross section! The cross section and its ratio to the \ttbar cross section of $0.024 \pm 0.009 (\mathrm{stat.}) \pm 0.001 (\mathrm{syst.})$ are in good agreement to the SM.

\section{Lorentz Invariance Violation (\dzero)}
\begin{figure}[ht]
     \centerline{
       \includegraphics[width=0.325\columnwidth]{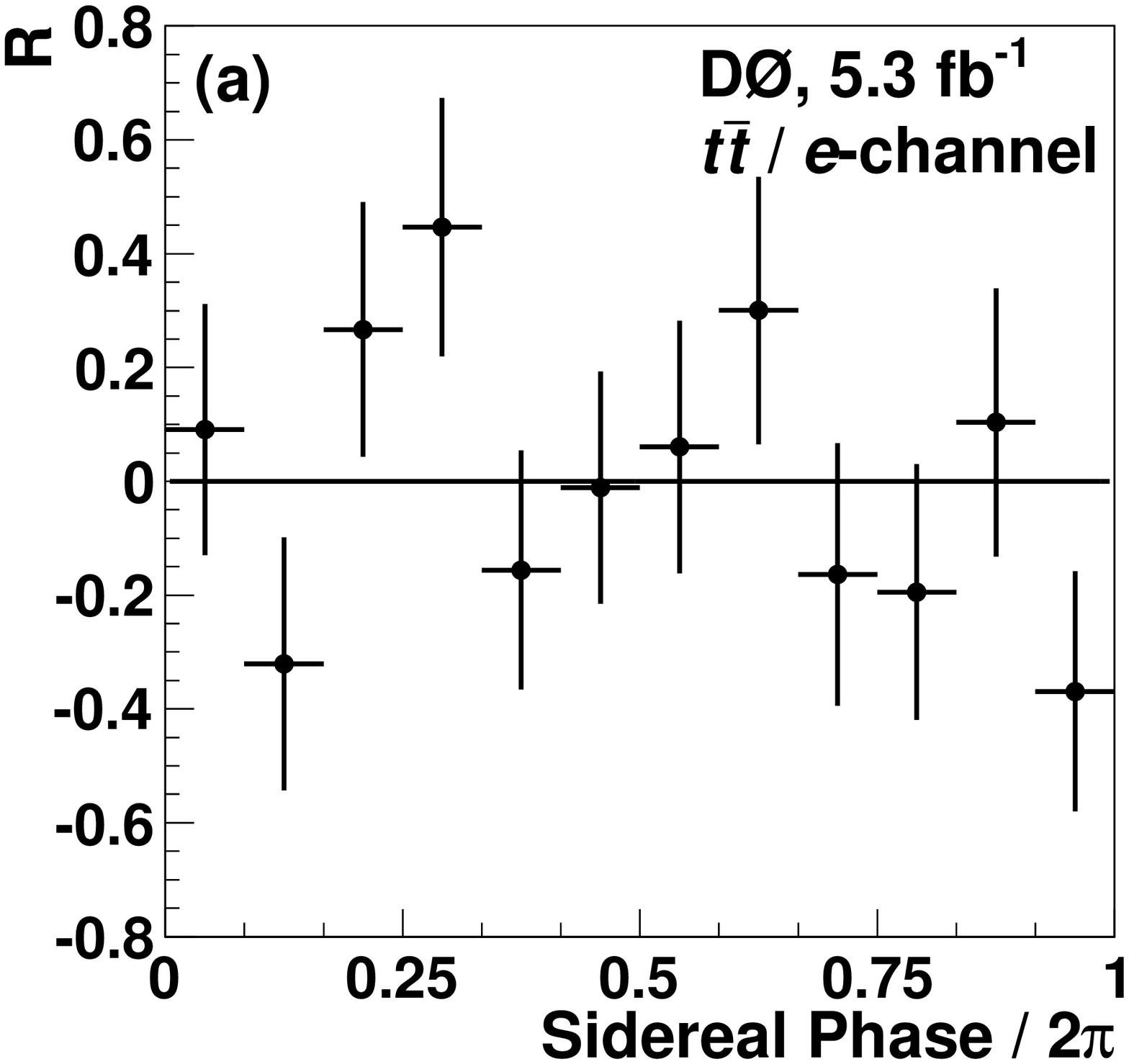}
       \includegraphics[width=0.325\columnwidth]{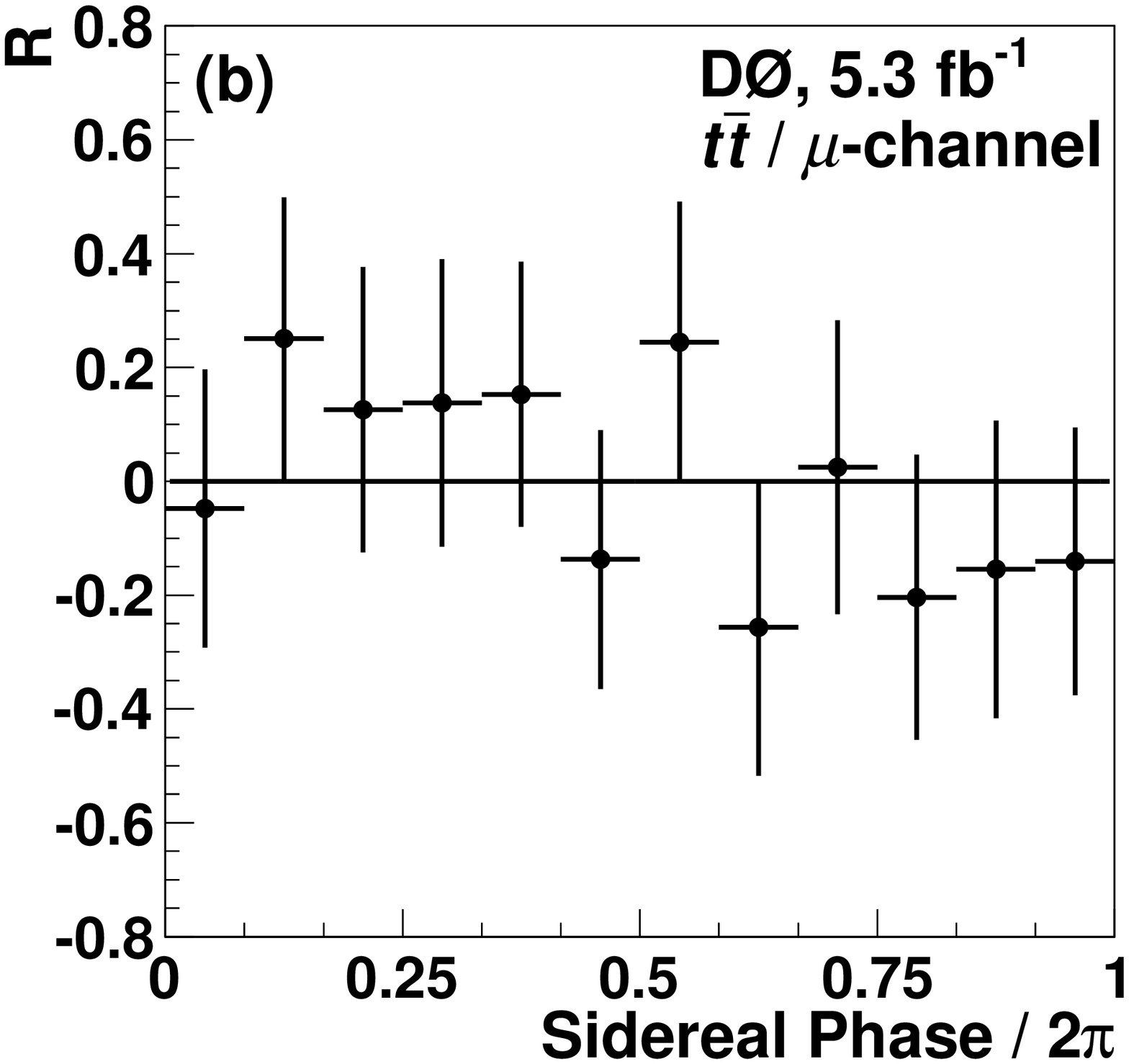}
     }
  \caption{\label{fig:livratio} (a) shows the background and luminosity corrected ratio $R$ as a function of the siderial phase (one siderial day) for events containing electrons, whereas (b) shows the same ratio $R$ for the muon case.}
 \end{figure}

\dzero searches for a time dependent \ttbar production cross section employing $5.3~\mathrm{fb^{-1}}$ of data \cite{LIVnote}. For the analysis \ttbar events in the \ljets~final state are selected with a lepton $(e/\mu)$, at least four jets, exactly one jet identified as a $b$-jet and \met. In addition the analysis relies on the timestamp of the data at production time. The Standard Model Extension (SME) \cite{SMEtheory} is an effective field theory and implements terms that violate Lorentz and CPT invariance. The modified SME matrix element adds Lorentz invariance violating terms for the production and decay of \ttbar events to the Standard Model terms. The SME predicts a cross section dependency on siderial time as the orientation of the detector changes with the rotation of the earth relative to the fixed stars. The luminosity-corrected relative \ttbar event rate ($R$) is expected to be flat within the Standard Model, i.e. no time dependency of the \ttbar production cross section. Figure \ref{fig:livratio} shows this ratio as a function of the siderial phase, \mbox{i.e.~1} corresponds to one siderial day.
There is no indication of a time dependent \ttbar production cross section. Instead this measurement sets the first constraints on Lorentz invariance violation in the top quark sector. As the top quark decays before it can hadronize the constraints are also the first ones for a bare quark.

\section{Measurement of the top-quark width (\dzero)}
An indirect determination of the top quark width at \dzero \cite{topwidth} is presented using the results of earlier measurements at \dzero. The total width $\Gamma_t$ is determined from the ratio of the partial decay with $\Gamma(t \rightarrow Wb) / {\cal B}(t \rightarrow Wb)$ under the assumption of the same $tWb$ couplings in production and decay. The partial decay width is derived from the single top $t$-channel cross section measurement of $2.90 \pm 0.59 (\mathrm{stat.+syst.})$ \cite{d0_singletop} corrected by the ratio of the SM expected values for partial decay width and single top $t$-channel cross section. To obtain the total decay width $\Gamma_t$ the derived partial decay width is corrected by the branching fraction ${\cal B}(t \rightarrow Wb) = 0.90 \pm 0.04$ mentioned also earlier in this text \cite{d0_R}.
\begin{figure}[h]
\includegraphics[width=0.50\columnwidth]{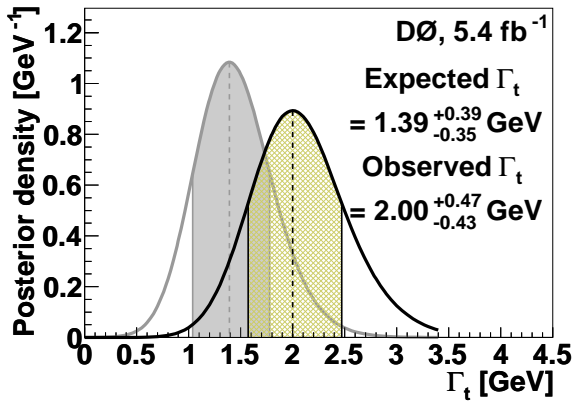}\hspace{2pc}%
\begin{minipage}[b]{14pc}\caption{\label{fig:topW} Probability density for the expected and observed total width $\Gamma_t$. The shaded areas represent
one standard deviation around the most probable value.}
\end{minipage}
\end{figure}
With $5.3~\mathrm{fb^{-1}}$ of data the currently most precise indirect determination of the top quark width results $\Gamma_t = 2.00 ^{+0.47}_{-0.43}$ GeV, which corresponds to a lifetime of $\tau_t = (3.29 ^{+0.90}_{-0.63}) \times × 10^{−25}$ s. Figure \ref{fig:topW} shows the probability density for the expected and measured total width $\Gamma_t$. As the partial decay width is determined using SM predicted values any deviation from the SM expected value of the total decay width would indicate contributions due to new physics. The result does not indicate any hints for new physics contributions.

\section{Conclusion}
Various recent measurements of top quark properties at the Tevatron are discussed. The presented results are in good agreement with the Standard Model expectations and do not show any hints for new physics. More details and results are given at the \dzero and CDF webpage \cite{webpages}.  CDF and \dzero continue to provide unique results in the top sector and more top quark measurements using the full data sample are expected to come out soon.

\ack
The author thanks the organizers of the TOP 2012 workshop for the invitation and for the hospitality of the conference venue.


\end{document}